\documentclass[a4paper,fleqn,usenatbib,useAMS]{mnras}

\usepackage{color}
\usepackage{graphicx}
\usepackage{amsmath}	
\usepackage{amssymb}	
\usepackage{multicol}        
\usepackage{bm}		
\usepackage{pdflscape}	
\usepackage{subcaption}
\usepackage{comment}

\usepackage[T1]{fontenc}
\usepackage{ae,aecompl}
\usepackage{enumitem}
\usepackage{colortbl}

\usepackage{newtxtext,newtxmath}

\usepackage{etoolbox}

\usepackage[dvipsnames]{xcolor}
\usepackage[colorinlistoftodos]{todonotes}
\definecolor{salmon}{rgb}{0.95,0.5,0.25}

\title{No globular cluster progenitors in Milky Way satellite galaxies}

\author[P. Boldrini and J. Bovy]{Pierre Boldrini$^{1,3}$\thanks{Contact e-mail: \href{mailto:boldrini@iap.fr}{boldrini@iap.fr}} and {Jo Bovy$^{2}$}
\\
$^{1}$Sorbonne Universit\'e, CNRS, UMR 7095, Institut d'Astrophysique de Paris, 98 bis bd Arago, 75014 Paris, France \\
$^{2}$Department of Astronomy and Astrophysics, University of Toronto, 50 St George Street, Toronto, ON M5S 3H4, Canada \\
$^{3}$ Universit\'e de Lorraine, CNRS, Inria, LORIA, F-54000 Nancy, France
}

\pubyear{2021}

\begin{document}
\label{firstpage}
\pagerange{\pageref{firstpage}--\pageref{lastpage}}
\maketitle

\begin{abstract}
In order to find the possible progenitors of Milky Way globular clusters, we perform orbit integrations to track the orbits of 170 Galactic globular clusters and the eleven classical Milky Way satellite galaxies backwards in time for 11 Gyr in a Milky-Way-plus-satellites potential including the response of the MW to the infall of the Large Magellanic Cloud and the effect of dynamical friction on the satellites. To evaluate possible past associations, we devise a globular-cluster--satellite binding criterion based on the satellite's tidal radius and escape velocity and we test it on globular clusters associated with the Sagittarius dwarf and on dwarf galaxies associated with the Large Magellanic Cloud. For these, we successfully recover the dynamical associations highlighted by previous studies and we derive their time of accretion by the Galaxy by using Gaia EDR3 data. Assuming that Milky Way globular clusters are and have been free of dark matter and thus consist of stars alone, we demonstrate that none of the globular clusters show any clear association with the eight classical dwarf spheroidal galaxies even though a large fraction of them are believed to be accreted. This means that accreted globular clusters either came in as part of now-disrupted satellite galaxies or that globular clusters may have had dark matter halos in the past -- as suggested by the similar metallicity between globular clusters and dwarf galaxies. 
\end{abstract}





\section{Introduction}

The origin of compact stellar clusters, also known as globular clusters (GCs), is one of the open questions in modern astrophysics. Many years now since their first discovery, it is still unclear whether they formed within dark matter minihalos or as gravitationally bound clouds in the early universe \citep{1968ApJ...154..891P,2005ApJ...623..650K,2015MNRAS.454.1658K,1984ApJ...277..470P,2017MNRAS.471L..31P}. Despite ongoing debates about their origin \citep{2015MNRAS.454.1658K,2018RSPSA.47470616F}, these gravitationally-bound groupings of mainly old stars also have a similarly long history of being used as probes of the galaxy formation and assembly process, particularly in the Milky Way (MW) \citep{1978ApJ...225..357S,1999AJ....117.1792D,2006ARA&A..44..193B}.

It has long been recognized that the accretion of satellite galaxies has contributed to the growth of the MW \citep{1978ApJ...225..357S}and the Galaxy has accreted around one hundred satellite galaxies \citep{2016ARA&A..54..529B}. As a natural result of such events, GCs may have been accreted by the MW \citep{2009MNRAS.399.1275P}. Indeed, several authors have used metallicities and horizontal-branch morphologies to distinguish GC origin; those formed in satellite galaxies and those formed in situ within the Galaxy \citep{1999AAS...194.4002Z,1993AJ....105..971V,2004MNRAS.355..504M}.

Recently, full six-dimensional phase-space information for almost all of the Galactic GCs from the early third data release of the Gaia mission has offered unique insights into their dynamics \citep{2019MNRAS.484.2832V}. Thanks to these data, it is now possible to assess whether GCs formed in the MW or in a satellite galaxy that was later accreted.

It was found that 62 of the present-day MW GCs likely formed in the MW, $\sim$55–65 have an extragalactic origin by using their dynamical properties in combination with their metallicity and their age \citep{2019A&A...630L...4M}. The rest likely has a more heterogeneous origin as they are not clearly associated to stellar debris in MW. \citep{2019MNRAS.486.3180K,2019A&A...630L...4M}. It was established that the main contribution to the MW came from three major accretion events: Sagittarius \citep{1994Natur.370..194I,2010ApJ...718.1128L}, Gaia-Enceladus \citep{2018MNRAS.478..611B,2018Natur.563...85H}, and Sequoia \citep{2019MNRAS.488.1235M}. In fact, numerous pieces of evidence indicate that 35$\%$ of the clusters are possibly associated with these merger events based on Gaia DR2 data \citep{2019A&A...630L...4M} or are formed in accreted dwarf galaxies based on HST photometry measurements \citep{2013MNRAS.436..122L}. 

A particularly intriguing set of GCs is the population of eleven loosely bound GCs or at high energies in the MW as defined by \cite{2019A&A...630L...4M}. These GCs (AM1, Eridanus, Pyxis, Palomar 3, Palomar 4, Crater, NGC 6426, NGC 5694, NGC 6584, NGC 6934 and Palomar 14) do not seem to have formed in situ in the Galaxy \citep{2019A&A...630L...4M}. In the E-MOSAICS simulations, high-energy GCs are generally old and metal-poor and then it may not be possible to unambiguously determine their origin \citep{2020arXiv200300076P}. Some of the 11 GCs have previously been tentatively associated with MW satellite galaxies such as Fornax, Sculptor, or the Large Magellanic Cloud \citep{1995ApJ...453L..21I,2014MNRAS.442L..85K}. Intriguingly, Fornax is the only one of the classical dwarfs to have six GCs in orbit \citep{1991AJ....102..914M,2003MNRAS.340..175M,2019ApJ...875L..13W}, thus demonstrating that dwarfs of the MW can have hosted GCs in the past that now have been stripped. Indeed, as the 11 loosely-bound GCs reach far distances from the MW centre over their history, they could have interacted with or belonged to MW satellites.

In this paper, we examine if the eleven satellite galaxies of the MW could be progenitors of some of the 170 GCs. We follow the dynamical history backwards in time of these GCs in the MW+satellite environment by using orbital integration methods. The paper is organized as follows. Section 2 provides a description of our orbital-integration method. In Section 3, we discuss how we decide whether or not a GC was associated with a satellite galaxy in the past and present our results on associations between GCs and the Sagittarius dwarf, satellites of the Large Magellanic Cloud and the Cloud itself, and finally all Milky-Way GCs and the classical satellites. Section 4 presents our conclusions.

\section{Method}\label{mwc}

In order to find the possible progenitors of the 170 Galactic GCs, we take their present-day phase-space position derived from Gaia EDR3 data \citep{2019MNRAS.484.2832V} and integrate them backwards in time for 11 Gyr in a model for the Milky Way's gravitational potential. It was recently demonstrated that interactions with MW satellites can cause the orbital properties of Galactic GCs to evolve significantly over time \citep{2020MNRAS.499..804G}. Therefore, we include the gravitational potential of the eleven classical satellites in the potential model (see Table~\ref{tab1}). Orbit integrations are performed with a time step of 10 Myr using the publicly available code \texttt{galpy}\footnote{Available at \url{https://github.com/jobovy/galpy}} \citep{2015ApJS..216...29B}
.

We assume that the main component of the Galactic potential is well represented by the \texttt{MWPotential2014} mass model from \cite{2015ApJS..216...29B}. Much of the stellar halo was likely accreted between 9 and 11 Gyr ago \citep{2019A&A...632A...4D,2020MNRAS.492.3631M}. Moreover, cosmological simulations of the formation of Milky-Way-like galaxies show that the main halo mass, which grows with time, reaches its asymptotic value 9-10 Gyr ago \citep{2007ApJ...657..262D,2010MNRAS.406.2312L}. As a consequence, we establish that our model describes the Galaxy well until 9-10 Gyr ago. 

For the satellites, we assume a \cite{1990ApJ...356..359H} density profile for the DM component. Given the halo mass and redshift, both halo concentrations $c_{200}$ can be estimated from cosmological $N$-body simulations \citep{2014MNRAS.441.3359D}. We neglect the stellar contribution of satellites as their dynamics is largely governed by the dark component. The mass loss histories of satellite galaxies are poorly constrained, but we can write down a simple mass-loss model by connecting the current mass $M_{\mathrm{200}}$ and the satellite mass $m(t)$ as: 
\begin{equation}
    m(t)=M_{\mathrm{200}}\left(1 - \frac{t}{T_{\rm d}}\right)
\label{equa1}
\end{equation}
where t and $T_{\rm d}$ are the time and the characteristic time defined as $m(-t_{\rm infall}) = M_{\mathrm{pi}}$ ,where $M_{\mathrm{pi}}$ is the pre-infall mass, respectively. We assume that the satellite mass remains constant before the infall time constrained by \cite{2019arXiv190604180F} (see Table~\ref{tab1}). However, incorporating this mass-loss model during all orbital integrations is computationally costly. As an alternative, we choose to assess the impact of mass loss by keeping the satellite masses constant during integrations and considering minimum and maximum values for the mass. They correspond to the current mass $M_{\mathrm{200}}$ (as the minimum mass M$_{\mathrm{min}}$) and the pre-infall mass $M_{\mathrm{pi}}$ (as the maximum mass M$_{\mathrm{max}}$) determined by \cite{2019MNRAS.487.5799R} and \cite{2018MNRAS.481.5073E}, respectively (see Table~\ref{tab1}). Once they are falling into the MW, the satellites loose mass due to tidal effects and dynamical friction. Taking into account the mass loss is crucial as it affects directly the orbital radius and hence the tidal radius. Below, we therefore check that satellite orbits for these mass limits cover the same region as if we have considered a mass loss for all satellites. Table~\ref{tab1} summarizes all the satellite properties adopted in the paper. We create the potential of the moving satellites by integrating their orbits backwards in time for 11 Gyr in \texttt{MWPotential2014} by applying dynamical friction to them with a constant mass of M$_{\mathrm{min}}$ or M$_{\mathrm{max}}$, given their current positions and proper motions from \cite{2018A&A...619A.103F}. Dynamical friction is implemented by following the semi-analytic model of \cite{2015MNRAS.454.3778P}. The GCs are then integrated in the combined \texttt{MWPotential2014} plus the time-dependent potential of each of the 11 classical satellites and dynamical friction on the GC itself.

Moreover, the massive LMC, with a mass of about $10^{11}$ M$_{\sun}$, can significantly perturb the gravitational potential of the MW \citep{2015ApJ...802..128G}. As demonstrated very recently by \cite{2021arXiv210608819B}, the orbits of MW dwarfs can be perturbed by the LMC infall via the implementation of perturbed MW potential modelled by \cite{2021MNRAS.501.2279V}. That is why, we also include the response of the MW to the LMC infall in all our orbital integrations. Specifically, we include the effect of the acceleration of the Milky Way frame using \texttt{galpy}'s support for including fictitious forces  arising from an accelerating reference frame (new in version \texttt{1.7.2}). To compute the Milky Way frame's acceleration, we first integrate the LMC's orbit in the \texttt{MWPotential2014} potential plus dynamical friction and then compute the induced acceleration of the Milky Way center's from the orbiting LMC. This approximate method for determining the frame acceleration matches the calculation from \cite{2021MNRAS.501.2279V} well.

A mass of $6\times10^6$ M$_{\sun}$ corresponds to the upper limit of the stellar mass of Galactic GCs 12 Gyr ago \citep{2019MNRAS.482.5138B}. The time to change the GC apocentre substantially due to dynamical friction is  $\approx M_{\mathrm{enclosed}}/M_{\mathrm{GC}}\times t_{\mathrm{dyn}}$ where $M_{\mathrm{enclosed}}$ is the host galaxy mass enclosed within the orbit and $t_{\mathrm{dyn}}$ is the orbital time \citep{2008gady.book.....B}. As the mass ratio between MW enclosed mass and GC masses is large, dynamical friction is inefficient over our timescale (11 Gyr). We thus neglect the mass loss to the clusters assuming that the GC mass remains constant during integrations. 

As proper motion uncertainties and measurement errors could affect the results and thus also the possible close encounters between GCs and satellite galaxies, our analysis takes account of uncertainties and measurement errors as follows. In order to establish associations between MW satellites and GCs, we Monte Carlo sample their 6D phase-space position by sampling the distance, the radial velocity and proper motions including the covariance using Gaussian error distributions with means and standard deviations given by \citet{2019MNRAS.484.2832V} for the GCs and by \citet{2018A&A...619A.103F} for the satellite galaxies. We neglect the uncertainty in the position on the sky. We then evaluate the association criterion described below for all Monte Carlo samples.

\begin{table}
\begin{center}
\label{tab:landscape}
\begin{tabular}{cccccccccccc}
 \hline
 dSph & M$_{\mathrm{min}}$ & M$_{\mathrm{max}}$ & t$_{\rm infall}$\\
  & [$10^{9}$ M$_{\sun}$] & [$10^{9}$ M$_{\sun}$] & [Gyr]\\
  \hline
  LMC & 100 & 198.8 & 4.0$^{a}$\\
  SMC & 26 & 73.9 & 4.0$^{a}$\\
  Sgr & 14 & 62 & 7.48$^{b}$\\
  UMi & 0.79 &  2.8 & 10.7$^{c}$\\
  Draco & 3.16 & 3.5 & 10.4$^{c}$\\
  Sculptor & 1.99 & 5.7 & 9.9$^{c}$ \\
  Sextans & 0.316 & 2.0 & 8.4$^{c}$ \\
  Leo I & 1.99 & 5.6 & 2.3$^{c}$\\
  Leo II & 0.316 & 1.6 & 7.8$^{c}$\\
  Carina & 0.398 & 0.8 & 9.9$^{c}$\\
  Fornax & 0.79 & 21.9 & 10.7$^{c}$\\
    \hline
\end{tabular}
\caption{{\it MW satellite properties:} From left to right, the columns give for each classical satellites: the current mass $M_{\mathrm{200}}$ \protect\citep{2018MNRAS.481.5073E} and the pre-infall mass $M_{\mathrm{pi}}$ \citep{2019MNRAS.487.5799R}; the infall time from: (a) \protect\cite{2012MNRAS.425..231R}, (b) \protect\cite{2017ApJ...847...42D} and (c) \protect\citep{2019arXiv190604180F}.}
\label{tab1}
\end{center}
\end{table}

\section{Result}

\begin{table*}
\begin{center}
\label{tab:landscape}
\begin{tabular}{ccccccccccccc}
 \hline
 Globular cluster & $p_{\rm Sgr}$(M$_{\mathrm{min}}$) & t$_{\mathrm{acc}}$(M$_{\mathrm{min}}$) & $p_{\rm Sgr}$(M$_{\mathrm{max}}$) & t$_{\mathrm{acc}}$(M$_{\mathrm{max}}$) & Previously associated\\
  &  & [Gyr] &  & [Gyr]\\
  \hline
  \hline
  likely associated with Sgr & \\
  Whiting 1 & 0.68 & -0.26 & 0.84 & -0.21 & Yes\\
  Terzan 7 & 0.94 & 0.0 & 0.99 & 0.0 & Yes\\
  Arp 2 & 0.96 & 0.0 & 0.98 & 0.0 & Yes\\
  Pal 12 & 0.66 & -0.19 & 0.93 & -0.15 & Yes\\
  Terzan 8 & 0.99 & 0.0 & 0.99 & 0.0 & Yes\\
  NGC 6715 & 0.99 & 0.0 & 0.99 & 0.0 & Yes\\
  \hline
  not associated with Sgr & \\
  NGC 5694 & 0.004 & - & 0.0 & - & No\\
  NGC 7006 & 0.02 & - & 0.0 & - & No\\
  Pal 4 & 0.009 & - & 0.0 & - & No\\
  Crater & 0.004 & - & 0.0 & - & No\\
  Pal 15 & 0.004 & - & 0.006 & - & No\\
  NGC 5904 & 0.014 & - & 0.032 & - & No\\
  NGC 288 & 0.0 & - & 0.08 & - & No\\
  NGC 2298 & 0.0 & - & 0.2 & - & No\\
    \hline
\end{tabular}
\caption{{\it Summary of results for the Sagittarius dwarf:} From left to right, the columns give for each GCs: the probability of having been bound to the Sgr assuming M$_{\mathrm{min}}$ and M$_{\mathrm{max}}$ for the 11 satellite galaxies, with their respective times of the accretion by the MW if GCs were previously associated sith Sgr dwarf according to \protect\cite{2010ApJ...718.1128L,2017A&A...598L...9M,2018ApJ...862...52S,2020A&A...636A.107B}. Only GCs with a non-zero probability for one of the dwarf masses are listed. We found that 6 GCs were effectively associated with the Sgr dwarf galaxy. NGC6715, Arp 2, Terzan 7 and Terzan 8 still belong to Sgr, whereas Whiting 1 and Palomar 12 were accreted less than 0.3 Gyr ago by the MW galaxy. According to our results, the other 145 GCs have never been associated with the Sgr dwarf.}
\label{tab2}
\end{center}
\end{table*}

\subsection{Globular-cluster--satellite association criterion}

In order to identify which GCs were associated with the eleven satellite galaxies as likely progenitors, we define two simple criteria based on the tidal radius and the escape velocity of satellites. 

The tidal radius is the distance to the last closed zero-velocity surface surrounding a galaxy. Since the final closed surface passes through a minimum of the combined effective potential of the satellite and the Galaxy $\Phi_\mathrm{eff}$, the distance to this surface can be determined with the condition: 
\begin{equation}
   \frac{d\Phi_\mathrm{eff}}{dx}(r_{\mathrm{t}},0,0)=0, 
\end{equation}
where the x-axis is taken to point away from the center of the Galaxy. Using the orbital radius $R_0$ and the mass $M_{\mathrm{SG}}$, we calculate the theoretical tidal radius of each satellite galaxies (SG) over 11 Gyr as derived by \cite{2008ApJ...689.1005B} as
\begin{equation}
   r_{\mathrm{t}}^{\mathrm{SG}}=\left(\frac{GM_{\mathrm{SG}}}{\Omega^2 \upsilon}\right)^{1/3},
   \label{eq2}
\end{equation}
where the orbital frequency $\Omega$ at $R_0$, the epicyclic frequency $\kappa$ at $R_0$, and a positive dimensionless coefficient $\upsilon$ are defined as:
\begin{align}
   \Omega^2 & =\left(d\Phi_{\mathrm{MW}}(R)/dR\right)_{R_0}/R_0\,,\\
   \kappa^2&=3\Omega^2 + \left(d^2\Phi_{\mathrm{MW}}(R)/dR^2\right)_{R_0}\,,\\
   \upsilon&=4-\kappa^2/\Omega^2\,.
\end{align}
Dynamical friction and mass-loss at pericentric passages act by reducing the satellite tidal radius forward in time, thereby increasing the probability of GC accretions by the MW. We, therefore, use the orbital radii of GCs and the satellite tidal radii to define a simple criterion to identify satellite galaxies as likely progenitors:
\begin{equation}
    r_{\mathrm{c}}^{\mathrm{GC}}(t_{\mathrm{i}}) < r_{\mathrm{t}}^{\mathrm{SG}}(t_{\mathrm{i}}),
\label{eq3}
\end{equation}
where $r_{\mathrm{c}}^{\mathrm{GC}}$ and $t_{\mathrm{i}}$ is the GC orbital radius centered on a galaxy and a specific time, respectively. 

As the tidal radius criterion only considers an association in the position space of both galactic objects, it is insufficient to identify past bound associations. To do this, we need to establish an additional criterion based on the escape velocity of likely progenitors. The escape velocity is the maximum velocity that GCs could have if they had been bound to a MW satellite. As the next step, therefore, we calculate the escape velocity of the satellite $v_{\mathrm{esc}}$ and compare it with the GC relative velocity to its putative parent galaxy $v^{\mathrm{GC}}$:
\begin{equation}
    v^{\mathrm{GC}}(t_{\mathrm{i}}) < v_{\mathrm{esc}}^{\mathrm{SG}}(t_{\mathrm{i}}).
\label{eq4}
\end{equation}
This criterion and the escape velocity in it is evaluated at the minimum separation between the GC and the satellite galaxy. The above criteria give the information needed to decide whether GCs were bound or unbound to a satellite in the past. In order to determine the probability of a GC being bound to MW satellites, we count the fraction of the Monte Carlo samples that satisfy our criteria based on Equations~\eqref{eq3} and ~\eqref{eq4} at each time. We assume that the most recent time $t_{\mathrm{i}}$, which satisfies the previous conditions, corresponds to the time of the GC accretion by the MW, t$_{\mathrm{acc}}$. Before this time, the GC dynamics is no longer governed by the MW, but by the satellite galaxy.

\subsection{Sagittarius dwarf and its globular clusters}

\begin{figure}
\centering
\includegraphics[width=\hsize]{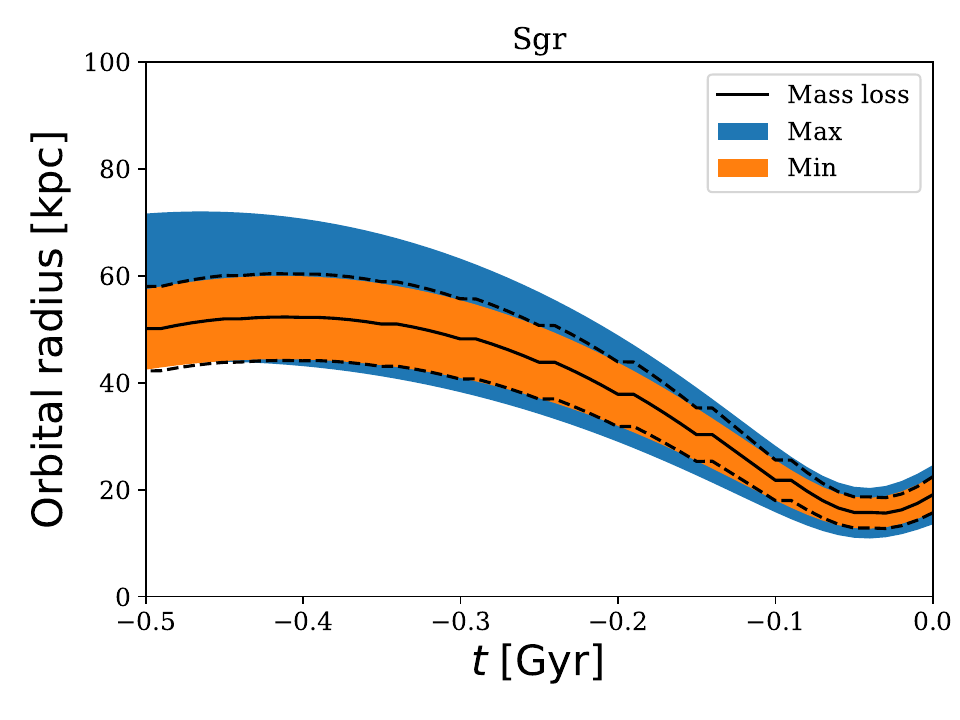}
\caption{\textit{Mass loss impact:} Orbital radius centered on the MW of the Sgr dwarf backwards in time over 0.5 Gyr in our galactic potential assuming M$_{\mathrm{min}}$, M$_{\mathrm{max}}$ and the mass loss model described by Equation~\eqref{equa1}. The bands correspond to the bound region around Sgr delimited by its tidal radius, which depends on its mass. Considering M$_{\mathrm{min}}$ and M$_{\mathrm{max}}$ masses for the orbital integrations is sufficient to approximate the mass loss of Sgr dwarf over 0.5 Gyr backwards in time.}
\label{fig1}
\end{figure}

First, we apply our method to the Sagittarius (Sgr) dwarf galaxy and the full set of 170 GCs. The ongoing disruption of the Sgr dwarf provides a formidable case study of GC accretions \citep{1994Natur.370..194I}. Six GCs such as NGC 6715, Ter 8, Ter 7, Arp 2, Pal 12 and Whiting 1 have been associated with this dwarf beyond any reasonable doubt \citep{2010ApJ...718.1128L,2017A&A...598L...9M,2018ApJ...862...52S,2020A&A...636A.107B}. However, there are still interesting candidates possibly associated with the Sgr stream such as NGC 2419, NGC 5634, NGC 4147 and NGC 5824 \citep{2020A&A...636A.107B,2019A&A...630L...4M}. It was also argued that NGC 2419 could be associated with Sgr dwarf in the past as this cluster is along the orbit of the Sgr stream \citep{2014MNRAS.437..116B,2018ApJ...862...52S,2017A&A...598L...9M,2020A&A...635L...3A}. We take the present-day phase-space position of all the 170 GCs derived from Gaia EDR3 data \citep{2019MNRAS.484.2832V} and integrate them backwards in time for 11 Gyr in our \texttt{MWPotential2014}+satellites potential including the response of the MW to the infall of the LMC. Table~\ref{tab2} shows the probability of having been bound to the Sgr dwarf, assuming M$_{\mathrm{min}}$ and M$_{\mathrm{max}}$ for the 11 satellite galaxies used in the time-dependent \texttt{MWPotential2014}+satellites potential, with the respective times of the accretion by the MW for the full set of 170 GCs. Only GCs with a non-zero probability for one of the satellite masses are listed. We found that 6 GCs were effectively associated with the Sgr dwarf galaxy. NGC6715, Arp 2, Terzan 7 and Terzan 8 still belong to Sgr, whereas Whiting 1 and Palomar 12 were accreted less than 0.3 Gyr ago by the MW galaxy, respectively (see Table~\ref{tab2}). According to our results, the other 164 GCs, including the recent candidates such as NGC 2419, NGC 5634, NGC 4147 and NGC 5824, have never been associated with the Sgr dwarf. Concerning NGC 2419, we stress that this GC is likely associated with a portion of the Sagittarius stream \citep{2017A&A...598L...9M,2018ApJ...862...52S,2020A&A...636A.107B}. That is why, it was pointed out that NGC 2419, should have been lost by Sgr dwarf more than 3-5 Gyr ago \citep{2020A&A...636A.107B}. This is consistent with the absence of dynamical association between NCG 2419 and the disrupted dwarf within the first 0.5 Gyr predicted by our method.

Figure~\ref{fig1} confirms that considering the M$_{\mathrm{min}}$ and M$_{\mathrm{max}}$ masses for the orbital integrations is sufficient to approximate the mass loss of Sgr over 0.5 Gyr backwards in time. Our method remains valid as long as the bound regions delimited by the satellite tidal radius for M$_{\mathrm{min}}$ and M$_{\mathrm{max}}$ cover the same region as if we have considered our mass loss model for Sgr over a period which includes the times of accretion that we found. Indeed, it is unlikely that their GCs got stripped very long in the past as Sgr is still disrupting and GCs are pretty tightly bound to their progenitor.

In light of previous studies, our results state that our method based on our criteria to associate GCs with Sgr as their progenitor works well and provides also an estimate of the time of accretion for GCs. 

\subsection{LMC and Dark Energy Survey dwarf galaxies}

\begin{figure}
\centering
\includegraphics[width=\hsize]{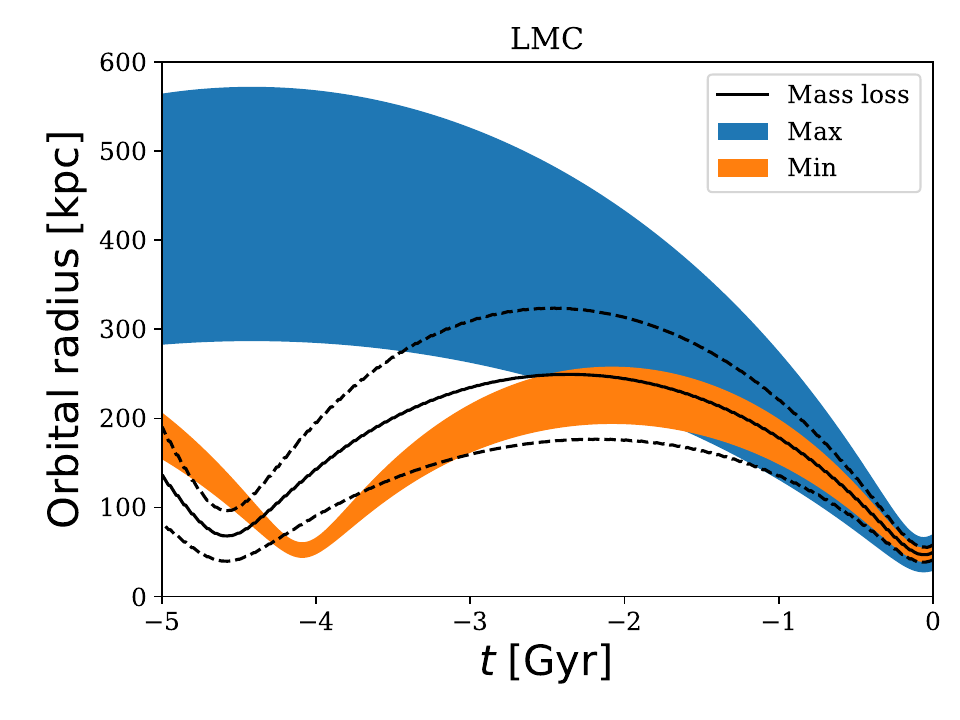}
\caption{\textit{Mass loss impact:} Orbital radius centered on the MW of the LMC backwards in time over 5 Gyr in our Galactic potential assuming M$_{\mathrm{min}}$, M$_{\mathrm{max}}$, and the mass loss model described by Equation~\eqref{equa1}. The bands correspond to the bound region around the LMC delimited by its tidal radius, which depends on its mass. Beyond 0.5 Gyr backwards in time, it is necessary to take correctly in account the mass loss for the LMC in order to find associations with DES dwarfs.}
\label{fig2}
\end{figure}

\begin{table*}
\begin{center}
\label{tab:landscape}
\begin{tabular}{ccccccccccccc}
 \hline
  DES dwarf & SMC & Tuc III & Car II & Phe II & Car III & Ret II & Aqu II & \\

 \hline 
     \hline
  $p_{\rm LMC}$ ($2\times10^{10}$ M$_{\sun}$) & 0.0 (0.0) & 0.0 (0.0) & 0.0 (0.0) & 0.0 (0.0) & 0.0 (0.0) & 0.0 (0.0) & 0.002 (0.003) \\
  t$_{\mathrm{acc}}$($2\times10^{10}$ M$_{\sun}$) & - & - & - & - & - & - & - \\
      \hline
  $p_{\rm LMC}$ ($5\times10^{10}$ M$_{\sun}$) & 0.0 (0.004) & 0.0 (0.002) & 0.0 (0.0) & 0.0 (0.0) & 0.0 (0.0) & 0.007 (0.007) & 0.008 (0.007)\\
  t$_{\mathrm{acc}}$($5\times10^{10}$ M$_{\sun}$) & - & - & - & - & - & - & - \\
      \hline
  $p_{\rm LMC}$ ($10\times10^{10}$ M$_{\sun}$) & 0.088 (0.08) & 0.042 (0.07) & 0.0 (0.0) & 0.003 (0.003) & 0.009 (0.016) & 0.11 (0.14) & 0.03 (0.027)\\ 
  t$_{\mathrm{acc}}$($10\times10^{10}$ M$_{\sun}$) & - & - & - & - & - & - & - \\
      \hline
  $p_{\rm LMC}$ ($15\times10^{10}$ M$_{\sun}$) & 0.38 (0.4) & 0.18 (0.22) & 0.0 (0.0) & 0.033 (0.026) & 0.002 (0.017) & 0.193 (0.024) & 0.04 (0.031) \\
  t$_{\mathrm{acc}}$($15\times10^{10}$ M$_{\sun}$) & - & - & - & - & - & - & - \\
      \hline
  $p_{\rm LMC}$ ($20\times10^{10}$ M$_{\sun}$) & {\bf 0.69 } ({\bf 0.68}) & 0.41 (0.5) & 0.0 (0.0) & 0.086 (0.076) & 0.027 (0.022) & 0.25 (0.21) & 0.031 (0.028)\\
  t$_{\mathrm{acc}}$($20\times10^{10}$ M$_{\sun}$) & -0.02 (-0.02) & - & - & - & - & - & - \\
      \hline
  $p_{\rm LMC}$ ($25\times10^{10}$ M$_{\sun}$) & {\bf 0.79} ({\bf 0.79}) & {\bf 0.61} ({\bf 0.69}) & 0.32 (0.3) & 0.13 (0.11) & 0.021 (0.026) & 0.25 (0.28) & 0.03 (0.025)  \\
  t$_{\mathrm{acc}}$($25\times10^{10}$ M$_{\sun}$) & -0.016 (-0.016) & -0.06 (-0.057) & - & - & - & - & - \\
    \hline
  $p_{\rm LMC}$ ($30\times10^{10}$ M$_{\sun}$) &  {\bf 0.79} ({\bf 0.81}) & {\bf 0.79} ({\bf 0.88}) & {\bf 0.99} ({\bf 0.99}) & 0.17 (0.17) & 0.016 (0.013) & 0.35 (0.034) & 0.009 (0.008) \\
  t$_{\mathrm{acc}}$($30\times10^{10}$ M$_{\sun}$) & -0.012 (-0.012) & -0.054 (-0.052) & 0.0 (0.0) & - & - & - & - \\ 
    \hline
  \cite{2020MNRAS.495.2554E}  & Yes & No & Yes & Yes & Yes & Yes & No  \\
 \cite{2020ApJ...893..121P}  & Yes & Yes & Yes & Yes & Yes & Yes & No \\
{\bf This work} & {\bf Yes} & {\bf Yes} & {\bf Yes} & No & No & No & No \\
    \hline
    \hline
\end{tabular}
\caption{{\it Summary of results for the LMC:} From top to bottow, the lines give for each DES dwarf: the probability of having been bound to the LMC $p_{\rm LMC}$ depending on its mass with their respective times of the accretion by the MW t$_{\mathrm{acc}}$; if these dwarfs were previously associated with LMC according to the litterature. All orbital integrations were performed in our MW+dwarf potential assuming M$_{\mathrm{max}}$ (M$_{\mathrm{min}}$) for the MW satellite galaxies including the response of the MW to the infall of the LMC. Only DES dwarfs with a non-zero probability are listed. We found that 3 DES dwarfs were effectively associated with the LMC. For a specific mass, we find that Car 2 still belongs to the LMC.}
\label{tab3}
\end{center}
\end{table*}

The Large Magellanic Cloud (LMC) is the most massive satellite of the MW and it is currently at a particular stage of its orbit. The LMC and the Small Magellanic Cloud (SMC) have been identified as a galaxy pair orbiting together in the MW \citep[e.g.][]{1990A&ARv...2...29W,2016ARA&A..54..363D,2006ApJ...652.1213K,2013ApJ...764..161K}. It has been suggested that the Clouds are just past their first closest approach to the Galaxy \citep{2007ApJ...668..949B,2011MNRAS.414.1560B,2017MNRAS.464.3825P}. The orbital behaviour depending on the LMC mass can be seen in Figure~\ref{fig2}. We see that the LMC is close to the pericentre of a highly eccentric orbit with long orbital times. Early studies have attempted to find tentative evidence of associated satellites with the LMC \citep[e.g.][]{2008ApJ...686L..61D,2011MNRAS.418..648S}.

Recently, the Dark Energy Survey (DES) mapped out a large fraction of the Southern Galactic hemisphere which uncovered 32 low-mass dwarf galaxies close  to  the  LMC \citep[e.g.][]{2015IAUGA..2256759K,2015ApJ...807...50B,2015ApJ...813..109D,2015APS..APR.X9002D,2015ApJ...808L..39K,2018MNRAS.479.5343K,2015ApJ...802L..18L,2016MNRAS.463..712T,2018MNRAS.475.5085T}. Given  the  proximity  of  the  DES survey area to the clouds, these new dwarfs appear good candidates to have been members of the LMC. One can easily imagine that some DES dwarfs have been stripped during the last passage of LMC, i.e., 0.14 Gyr ago. According to previous studies, only about six of the DES dwarfs are very likely to have been associated with the LMC in the past \citep{2016MNRAS.461.2212J,2021MNRAS.tmp.1022S,2017MNRAS.465.1879S,2018ApJ...867...19K,2019MNRAS.489.5348J,2020MNRAS.495.2554E,2020ApJ...893..121P,2021arXiv210608819B,2022MNRAS.511.2610C}.

To investigate possible associations of the DES dwarfs with the LMC, we apply our method to the 26 DES dwarfs considered by \cite{2020MNRAS.495.2554E}. We similarly Monte Carlo the 6d position and rewind these dwarfs and the LMC for 5 Gyr.
As in \cite{2020MNRAS.495.2554E}, we consider a grid of LMC masses (see Table~\ref{tab3}). This give us a range of tidal radii between 8 and 20 kpc. Our results suggest that the SMC, Tuc III, and Car II were effectively associated with the LMC. As in \cite{2020ApJ...893..121P}, we find that the 3 DES dwarfs were bound to the LMC in the last 0.5 Gyr. For a high LMC mass, we find that Car 2 still belongs to the LMC. In view of these recent accretions by the MW, it is reasonable to assume that dynamical friction is negligible on this timescale (see Table~\ref{tab3}). Besides, we stress that beyond 0.5 Gyr backwards in time, it is necessary to correctly take the mass loss for the LMC in account in order to find associations with DES dwarfs (see Figure~\ref{fig2}). Concerning the dynamical associations, Table~\ref{tab3} emphasises the impact of the LMC mass, which is still not sufficiently constrained, on the dynamical history of DES dwarfs. Very recently, \cite{2022MNRAS.511.2610C} investigate in the interval between -3 and -1 Gyr the past associations of the LMC with a fixed mass in a MW+LMC potential. However, current orbital methods need to take into the mass loss for the LMC in order to investigate old accretions by this satellite, i.e., 1-9 Gyr ago as depicted by Figure~\ref{fig2}.

In contrast to our results, previous studies based on orbital integrations found five other DES dwarfs that were likely associated with the LMC in a combined MW+LMC potential \citep{2020MNRAS.495.2554E,2020ApJ...893..121P}. One of the reasons that could explain this discrepancy is that we incorporate a potential of the moving satellites in addition the MW potential including the response of the MW to the infall of the LMC contrary to previous studies. Indeed the inclusion of MW dwarfs in our host potential can perturbed the orbits of DES dwarfs.  

Another reason is that, contrary to \cite{2020MNRAS.495.2554E} and \cite{2020ApJ...893..121P}, our MW satellites are modelled by a Hernquist profile with a tidal-radius criterion for boundedness. Indeed, our criterion based on the tidal radius is more restrictive when we have determined which galaxies are dynamically associated with the LMC. \citet{2020MNRAS.495.2554E} have only verified whether the energy relative to the LMC is less than the binding energy, while we take into account the influence of the MW on the LMC tidal radius. Even subject to this condition, a satellite could be unbound if it is orbiting outside of the tidal radius. \citet{2020ApJ...893..121P} have considered an outer radius for the LMC, which is assumed to be constant over time. In addition, this specific radius is significantly greater than our LMC tidal radii in the first Gyr. That is the reason why we expect that their criterion gives a higher number of MW satellites that have interacted with the LMC even if the MW response to the LMC is taking into account.

\subsection{Progenitors of 170 GCs}

\begin{figure*}
\centering
\includegraphics[width=0.7\hsize]{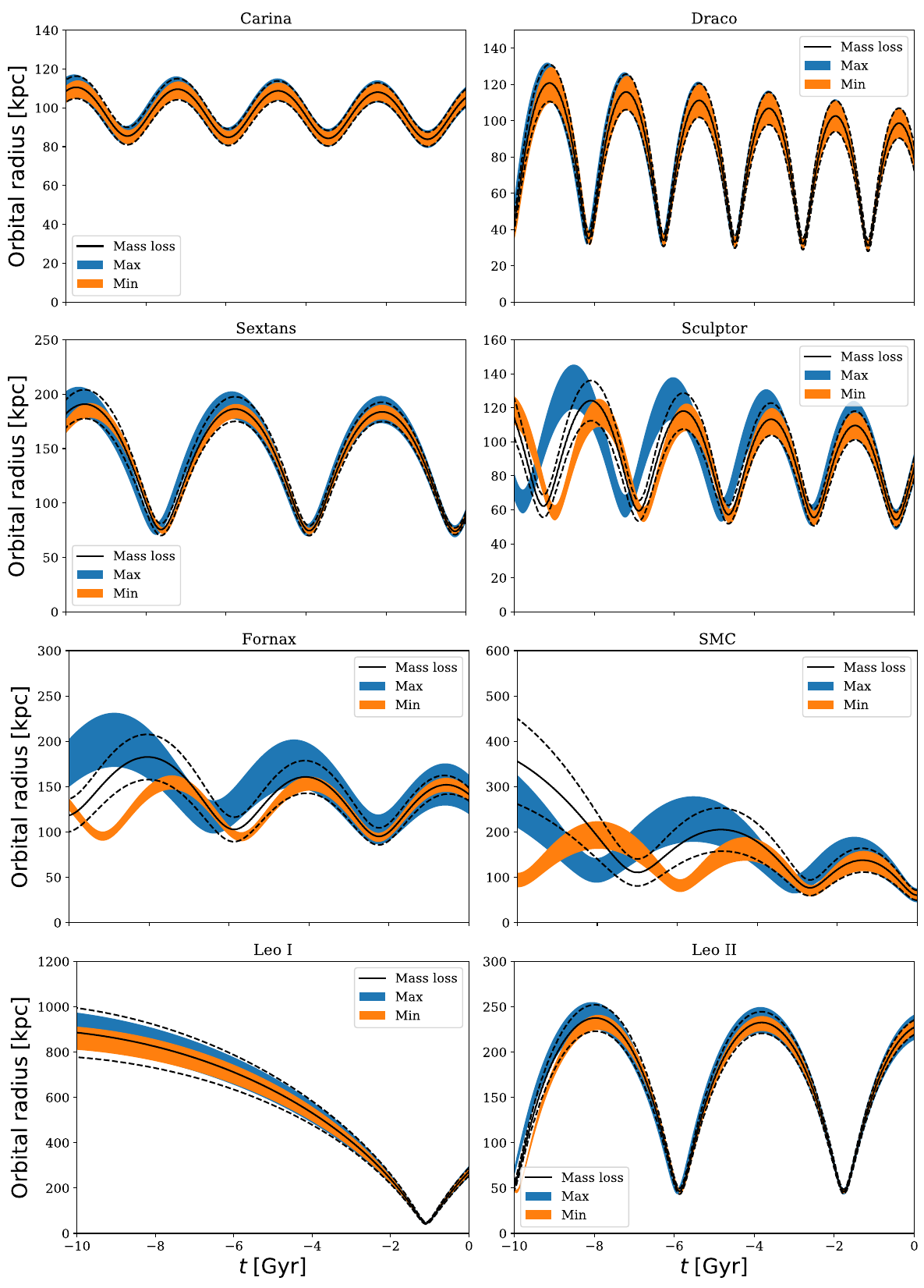}
\caption{\textit{Mass loss impact:} Orbital radius centered on the MW of the eight classical dwarfs backwards in time over 5 Gyr in our Galactic potential assuming M$_{\mathrm{min}}$, M$_{\mathrm{max}}$, and the mass loss model described by Equation~\eqref{equa1}. The bands correspond to the bound region around the dwarfs delimited by their tidal radius, which depends on their mass. Considering M$_{\mathrm{min}}$ and M$_{\mathrm{max}}$ masses for the orbital integrations is sufficient to approximate the mass loss of most of the classical dwarfs over 10 Gyr backwards in time. Beyond 6 Gyr backwards in time, it seems necessary to correctly take the mass loss for Fornax and the SMC into account in order to investigate associations with MW GCs.}
\label{fig3}
\end{figure*}

\begin{table}
\begin{center}
\label{tab:landscape}
\begin{tabular}{ccccccccccccc}
 \hline
 Progenitor & Globular cluster & $p_{\rm SG}$(M$_{\mathrm{min}}$) & $p_{\rm SG}$(M$_{\mathrm{max}}$) \\
   \hline
  \hline
  &  &  &  & \\
  Fornax & E1  & 0.005 & 0.05 & \\
  & Eridanus & 0.0 & 0.003 & \\
  & Pal 3 & 0.0 & 0.02 & \\
  & Crater & 0.02 & 0.08 & \\
  & Pal 4 & 0.0 & 0.006 & \\
  & Pal 14 & 0.03 & 0.001 & \\
  &  &  &  & \\
  Carina & E1 & 0.0 & 0.006 & \\
  & Eridanus & 0.001 & 0.0 & \\
  & Pal 3 & 0.0 & 0.004  & \\
  & Pal 4 & 0.0 & 0.006  & \\
  & Pal 14 & 0.0 & 0.005 & \\
  &  &  &  & \\
  Sculptor & E1 & 0.0 & 0.003 & \\
  & Pal 3 & 0.0 & 0.004 & \\
  &  &  &  & \\
  Umi & NGC 2419 & 0.002 & 0.004 & \\
  & Pal 4 & 0.001 & 0.0 & \\
  & Pal 14 & 0.002 & 0.0 & \\
  &  &  &  & \\
  Sextans & Pal 3 & 0.0 & 0.06 & \\
  &  &  &  & \\
  Draco & Pal 14 & 0.0 & 0.004 & \\
  &  &  &  & \\
  LMC & NGC 7006 & 0.02 & 0.08 & \\
  \hline
\end{tabular}
\caption{{\it Absence of GC progenitors in MW satellites:} From left to right, the columns give for each progenitor candidate: likely associated GCs; the probability of having been bound to the progenitor candidate assuming M$_{\mathrm{min}}$ and M$_{\mathrm{max}}$ for the MW satellite galaxies. Only GCs with a non-zero probability for one of the dwarf masses are listed. None of these GCs show any clear association with the eight classical MW dwarfs.}
\label{tab4}
\end{center}
\end{table}

By integrating orbits of the 170 GCs from Gaia EDR3 data in our MW+satellites potential including the response of the MW to the infall of the LMC, we apply our two dynamical criteria described by Equations~\eqref{eq3} and ~\eqref{eq4} in order to find likely past associations between theses GCs and MW satellites. Table~\ref{tab4} summarizes all the non-zero binding probabilities for possible satellite progenitors as a result of the investigation of associations between satellite galaxies and GCs over 10 Gyr backwards in time.

As for our previous tests applied on globular clusters and satellites associated with the Sagittarius dwarf and with the Large Magellanic Cloud, we stress that our two criteria do not lead to wrong associations. GCs, which are not listed in our Table~\ref{tab4}, have zero probability of having been bound in the past to one of the eight classical dwarfs. Besides, none of the GCs have a probability of association higher than $7\%$. Thus, we establish that none of these GCs show any clear association with the eight classical MW dwarf galaxies. More precisely, we find that none of 62 GCs likely formed in the MW was associated with MW satellites. 

Considering M$_{\mathrm{min}}$ and M$_{\mathrm{max}}$ masses for the orbital integrations is sufficient to approximate the mass loss of most of the classical dwarfs over 10 Gyr backwards in time according to Figure~\ref{fig3}. We emphasize again that our method remains valid as long as the bound regions delimited by the satellite tidal radius for M$_{\mathrm{min}}$ and M$_{\mathrm{max}}$ cover the same region as if we have considered our mass loss model (black dotted lines) for MW dwarfs, described by Equation~\eqref{equa1}. More precisely, it seems necessary to correctly take the mass loss for Fornax and the SMC into account in order to investigate associations with MW GCs beyond 6 Gyr backwards in time (Figure~\ref{fig3}).

Besides, Table~\ref{tab4} depicts that no GCs that belonged to the LMC even though the LMC is very massive. However, the LMC is orbiting at very large distance compared to MW GCs after only 1 Gyr backwards in time according to our Figure~\ref{fig2}. Indeed, most of MW GCs are currently located within 40 kpc from MW centre. In addition, the LMC only recently fell into the MW. Therefore, it is likely that if GCs were associated with the LMC, they were recently accreted by the MW. Besides, there are GCs still related to the LMC but there may not be any in our sample of 170 Gaia GCs. For instance, the eleven LMC GCs studied by \cite{2021NatAs...5.1247M} are not contained in our sample.

According to our results, none of the MW satellites investigated here appears to be progenitors of MW GCs. Some GCs and particularly AM1, Eridanus, Pyxis, Palomar 3, Palomar 4, Crater, NGC 6426, NGC 5694, NGC 6584, NGC 6934 and Palomar 14 do not seem to have formed in situ in the Galaxy. This population of eleven loosely bound GCs defined by \cite{2019A&A...630L...4M} support an accretion origin according to their dynamical properties of GCs found in the literature. Even though there is no observation of stellar debris from dwarf progenitors, there are several indications that Eridanus might be accreted such as that the probably orbit of Eridanus passes close to Fornax and Sculptor \citep{2017ApJ...840L..25M}. It was proposed that Pyxis probably originated in an unknown galaxy, which today is fully disrupted by the MW \citep{2017ApJ...840...30F}. Moreover, this cluster is likely to be associated with a narrow stellar stream crossing the constellations of Sculptor and Fornax \citep{1995ApJ...453L..21I,2014MNRAS.442L..85K}. Being born in a now detached or disrupted dwarf galaxy could explain the highly eccentric orbit of Palomar 4 \citep{2017MNRAS.467..758Z}. It was also suggest that Palomar 14 and NGC 5694 have dynamical properties in favor of an accretion origin \citep{2011ApJ...726...47S,2012A&A...537A..83C,2014MNRAS.443..815F,2006ApJ...646L.119L}. We underline that the hypothesis that these eleven GCs were stripped from dwarf galaxies is further supported by their metallicity indicating that these clusters are metal-poor as MW dwarfs \citep{2009A&A...506..729K,2010A&A...517A..59K,2009AJ....137.4361D,2011ApJ...738...74D,2010arXiv1012.3224H,1995AJ....109.1112S,1996AJ....112.1487H,1992AJ....104..164A,1999AJ....117..247S,2008AJ....136.1407D,2000PASP..112.1305P}.  

In this work, GCs were considered as purely stellar systems. Hovewer, it has also been proposed that GCs may have a galactic origin, where GCs are formed within small dark matter halos in the early Universe \citep[e.g.][]{2002ApJ...566L...1B,2005ApJ...619..243M,2005ApJ...619..258M,2016ApJ...831..204R,1984ApJ...277..470P,2021arXiv211201265V}. By adding this dark component, the dynamical friction acting on GCs with DM would no longer be negligible and GCs could then have been orbiting in the outer regions of the MW for a significant period in the past. This would allow the GCs to have interacted even more with the satellite galaxies, which populate the MW periphery. We leave a detailed investigation of this possibility to future work.

\section{Conclusion}

Using orbital integrations, we have tracked the orbits of the 170 Galactic GCs and the eleven satellite galaxies backwards in time over 11 Gyr in a MW + satellites potential, including the effect of dynamical friction on the satellites and the response of the MW to the infall of the LMC. In order to associate Galactic GCs and MW satellites, we apply a criterion based on the comparison between the GC orbits and the satellite tidal radius, which evolves with time. In addition, we establish a second criterion based on the escape velocity of likely progenitors to test for boundedness.

First of all, we have tested our method on the Sgr dwarf and its GCs. We successfully recovered the 6 GCs, which were previously associated with the Sgr dwarf beyond any reasonable doubt, without finding any spurious associations. For the 6 associated GCs, we also derived their corresponding time of accretion by the MW. Furthermore, we have also re-investigated the possible associations between DES dwarfs and the LMC. Orbital integrations allow us to successfully recover the recent dynamical associations, i.e in the last 0.5 Gyr, where the mass loss of the LMC is negligible.

Concerning the 170 GCs, we have looked for their likely dynamical associations with the MW satellites backwards in time for 11 Gyr in a Milky-Way-plus-satellites potential by including the MW response to the LMC infall. For this purpose, we estimate the probability of having been bound to the dwarf galaxies for all the GCs. Assuming that these clusters are and have been free of dark matter and thus consist of stars alone, we find that none of these GCs show any clear association with the eight classical MW dwarf galaxies. Indeed, we also retrieve that none of 62 GCs thought to have formed in the MW in the literature was associated with MW satellites.

Nevertheless, a large fraction of GCs are believed to be accreted, especially the eleven loosely bound GCs identified by \cite{2019A&A...630L...4M}. If these accreted GCs were formed in their own dark matter subhalos or in now-disrupted satellite galaxies, investigating their orbital history requires a much more complex modeling that will have to rely on $N$-body simulations. Another challenging problem is to fully and consistently model the MW's mass, which has drastically grown before $z=2$ due to mergers \citep{2017MNRAS.465.3622R}. This will be crucial for investigating GC accretions at this earlier epoch than we have considered in this paper.

\section{Acknowledgments}

We thank the reviewer for their constructive feedback which helped to improve the quality of the manuscript. We thank also David Valls-Gabaud and Joe Silk for illuminating discussions about globular clusters. JB acknowledges financial support from NSERC (funding reference number RGPIN-2020-04712) and an Ontario Early Researcher Award (ER16-12-061).

\section{Data availability}
This work has made use of data from the Milky-Way globular clusters \citep{2019MNRAS.484.2832V} and satellite galaxies \citep{2018A&A...619A.103F} catalogs with data from Gaia EDR3 \citep{2018A&A...616A..12G}.


\bsp	
\label{lastpage}
\end{document}